\documentclass[twocolumn,showpacs,preprintnumbers,amsmath,amssymb,aps,prb]{revtex4}
\usepackage{graphicx}% Include figure files

\usepackage{dcolumn}% Align table columns on decimal point
\usepackage{bm}% bold math
\usepackage{mathrsfs}
\usepackage{epsfig}
\usepackage{graphicx}
\usepackage{amsmath}
\usepackage{amsfonts}
\usepackage{amssymb}
\usepackage{mathrsfs}
\usepackage{multirow}

\begin{document}

\title{Collective vortex phases in periodic plus random pinning potential}
\author{W. V. Pogosov$^{1,2}$, V. R. Misko$^{1}$, H. J. Zhao$^{1}$, 
and F. M. Peeters$^{1}$} 
\affiliation{$^{1}$Departement Fysica, Universiteit Antwerpen, Groenenborgerlaan 171,
B-2020 Antwerpen, Belgium}
\affiliation{$^{2}$Institute for Theoretical and Applied Electrodynamics, Russian Academy
of Sciences, Izhorskaya 13, 125412, Moscow, Russia}
\date{\today }

\begin{abstract}
We study theoretically the simultaneous effect of a regular and a random
pinning potentials on the vortex lattice structure at filling factor one.
This structure is determined by a competition between the square symmetry of
regular pinning array, by the intervortex interaction favoring a triangular
symmetry, and by the randomness trying to depin vortices from their regular
positions. Both analytical and molecular-dynamics approaches are used. We
construct a phase diagram of the system in the plane of regular and random
pinning strengths, and determine typical vortex lattice defects appearing in
the system due to the disorder. We find that the total disordering of the
vortex lattice can occur either in one or in two steps. For instance, in the
limit of weak pinning, a square lattice of pinned vortices is destroyed in
two steps. First, elastic chains of depinned vortices appear in the film,
but the vortex lattice as a whole remains still pinned by the underlying
square array of regular pinning sites. These chains are composed into
fractal-like structures. In a second step, domains of totally depinned
vortices are generated and the vortex lattice depins from regular array.
\end{abstract}

\maketitle

%{Vortex lattices, flux pinning, flux creep}

\section{Introduction}

Magnetic and current-carrying properties of superconducting films with
nanoengineered arrays of periodic and quasi-periodic pinning sites attract a
lot of attention both from the experimental \cite{1,2,3,4,5,6,7,8,9,10,11}
and theoretical \cite{12,13,14,15,16,17,18} points of view. The main reason
of this interest is that the regularity in positions of pinning sites
produces a large increase of the critical current at certain values of the
magnetic field which correspond to the matching between the number of
vortices and the number of pinning sites in the system. The highest value of
the critical current in films with periodic pins was obtained for the same
concentration of vortices and pinning sites. If the pinning potential is
strong enough, vortices in a film with a square array of pinning sites also
form a square lattice instead of the triangular one, the latter being
energetically favorable in the absence of pinning. This regime corresponds
to single-vortex pinning when vortices are pinned individually and the
flux-line lattice behaves not as a collective media. However, if the pinning
strength becomes of the order of the elastic energy of the flux-line
lattice, the vortex-vortex interaction becomes important, and the triangular
symmetry of the vortex array is recovered. Thus, at the filling factor 1,
low values of the pinning strengths favor a deformed triangular vortex
lattice, which is basically depinned from the pinning array. At high values
of the pinning strength, the regular square array of pinned vortices is the
lowest energy state. There is also an intermediate regime, where the vortex
lattice is regular, but half of the vortex rows is depinned and form a
lattice with a symmetry close to the triangular one. These three vortex
phases are depicted in Fig. 1.

Structural phase transitions due to the tuning of the regular pinning
strengths were studied theoretically within the framework of the
London theory \cite{Ukr,Reich2001,Pogosov} where vortices were treated as
point-like objects, i.e., near the first critical field. The same regime was
considered in Refs. \cite{VVM,Reich} by using molecular dynamics
simulations. Ref. \cite{Maniv} analyzes such transitions near the second
critical field using the linearized Ginzburg-Landau theory, whereas in Refs. 
\cite{16,17,Golib} the whole system of the Ginzburg-Landau equations was
solved numerically. There are also applications of the same ideas for the
vortex lattices in rotating Bose-Einstein condensates of alkali metal atoms,
where regular pinning potential can be formed and easily tuned, in a broad
range, by using lasers. At first, structural transitions of vortex lattices
in this system were predicted theoretically in Refs. \cite{R1,R2,Bigelow}
using Gross-Pitaevskii equation and then observed experimentally \cite%
{Cornell}. We also would like to mention a recent experimental work \cite%
{French} on elastic lattices of millimeter-size charged particles in a
square array of traps which exhibit similar physics as the here considered
vortex phases. Another example is colloidal particles on periodic substrates 
\cite{colloids1,colloids2}. Thus, the problem of the competition between the
symmetry of the underlying pinning array and the lattice of repealing
particles is important in various fields of modern physics. Note that in the
case of superconducting films, holes which are usually used to create
regular pins produce a strong pinning potential, which always dominate the
vortex-vortex interaction. Weak enough pinning sites can be fabricated, for
example, by imposing a periodic modulation on the film surface and were
recently studied theoretically in Ref. \cite{Golib}.

In the present paper, we consider theoretically a more complicated
situation, when there is an additional factor, namely, a random pinning
potential. Random pinning is always found in any real physical system;
moreover randomness cannot be ruled out even in numerical simulations, where
its effect should be carefully analyzed. Samples potentially important for
applications have to be very large compared to the period of the regular
pinning potential, and therefore disorder and defects in lattices of
vortices or other interacting objects are unavoidable. For this reason,
understanding of basic properties of such systems is of scientific and
hopefully of practical importance.

Thus the structure of the vortex lattice in the system under consideration
is determined by three factors: (i) a square array of regular pinning sites,
which tries to impose its own symmetry on the vortex lattice, (ii) an
interaction between vortices, which favors a triangular configuration of the
vortex lattice, (iii) a random pinning potential, which attempts to destroy
the regularity in the vortex positions and to depin them from the square
lattice. We analyze here analytically various kinds of vortex lattice
defects for a two-dimensional system and determine typical values of the
random pinning strengths leading to a spontaneous generation of these
defects. This allows us to construct an approximate phase diagram of the
system in the plane of regular pinning strength and random pinning strength.
We also perform molecular-dynamic simulations in order to model numerically
the process of disordering of the square vortex array. The obtained vortex
patterns and the typical values of random pinning strength producing
disordering are in good agreement with our expectations obtained from the
analytical treatment.

Note that the problem studied can be also linked to other situations when
there is a competition of randomness and regularity, but where the
dimensionalities of the pinning potential and an elastic media are possibly
different from two-dimensional. These are flux-line lattices in layered
superconductors, where layers act as one-dimensional pinning centers giving
rise to a so-called intrinsic pinning. Such pinning centers can also be
created by twin boundaries. Since there is always disorder in real
superconductors, one can also have a competition between randomness and
regularity \cite{Zhukov,AL,Fisher}. Other examples are charge and spin
density waves.

The paper is organized as follows. In Section II, we present the basic
formulation of our model and discuss our approach of molecular dynamic
simulations. In Section III, various kinds of kinks or defects in vortex
lattices are found, their structures are discussed, and their energies and
sizes are estimated. In Section IV, we characterize a random potential and
explain how to account its effect on vortex lattices. Our main results are
given in Section V, where we construct an approximate phase diagram of the
system in the plane of regular and random pinning strengths. Finally, we
conclude in Section VI.

\section{Model}

\subsection{Basic formalism}

In our model, we treat vortices as point-like objects interacting via a
pair-wise potential, which are valid assumptions as long as the applied
magnetic field is much lower than the second critical field $H_{c2}$, and
the Ginzburg Landau parameter is large, $\kappa \gg 1$. Under such
assumptions, the interaction energy of two vortices positioned at $\mathbf{r}%
_{1}$ and $\mathbf{r}_{2}$\ can be well approximated by the well-known
London expression

\begin{equation}
H_{int}(\mathbf{r}_{1},\mathbf{r}_{2})=\frac{2\pi }{\kappa ^{2}}K_{0}\left(
\left\vert \mathbf{r}_{1}-\mathbf{r}_{2}\right\vert \right) ,  \tag{1}
\end{equation}%
where $K_{0}$ is a modified Bessel function. Here and below the following
dimensionless variables are used: distances are measured in terms of the
London penetration depth $\lambda $ and energy is measured in units of $%
H_{c}^{2}\lambda ^{3}/8\pi $ with $H_{c}$\ being the thermodynamic critical
field. An important quantity, which we are going to use, is the interaction
energy of a regular vortex row with a given vortex situated outside of this
row. If we put a center of coordinates at one of the vortices in the row,
and the $y$-axis is along the row, then the interaction energy of the vortex
located at ($x$, $y$) is given by

\begin{equation}
H_{int}^{row}(x,y)=\frac{2\pi }{\kappa ^{2}}\sum_{m=-\infty
}^{\infty }K_{0}\left( \sqrt{x^{2}+(y+md)^{2}}\right) ,  \tag{2}
\end{equation}%
where $d$\ is the intervortex distance. Using Fourier transformation for a
modified Bessel function $K_{0}(r)$ and performing a summation in the
reciprocal space, Eq. (2) can be rewritten as%
\begin{equation}
H_{int}^{(row)}(x,y) = \frac{2\pi }{\kappa ^{2}}\frac{\pi }{d}
\sum_{m=-\infty }^{\infty }\frac{1}{\sqrt{1+\frac{4\pi ^{2}m^{2}}{d^{2}}}} 
\tag{3}
\end{equation}
\begin{equation*}
\times \exp \left( -x\sqrt{1+\frac{4\pi ^{2}m^{2}}{d^{2}}}\right) \cos \frac{%
2\pi my}{d}.
\end{equation*}

Vortices are located in a two-dimensional film with a periodic square array
of pinning sites. To describe the pinning potential of a single site, we use
a parabolic function:%
\begin{equation}
V(\mathbf{r})=-U_{0}\left( 1-\left( \frac{\left\vert \mathbf{r}\right\vert }{%
\sigma }\right) ^{2}\right)  \tag{4}  \label{pinpot}
\end{equation}%
inside the well, $r\leq \sigma $, and the potential is zero outside the
well, $r>\sigma $. In this paper, we consider the situation when the well
radius $\sigma $ is much smaller than the intersite distance $a$, i.e., $%
\sigma \ll a$.

In addition to the regular pinning potential there is also a random pinning
potential in the system; we denote its energy by $\varepsilon (\mathbf{r})$.
The total energy of the system is thus given by

\begin{equation}
E=\sum_{i,j}H_{int}(\mathbf{r}_{i},\mathbf{r}_{j})+\sum_{i,n}V(\mathbf{R}%
_{n}-\mathbf{r}_{i})+\sum_{i}\varepsilon (\mathbf{r}_{i}),  \tag{5}
\end{equation}%
where $\mathbf{r}_{i}$ and $\mathbf{R}_{n}$ stand for the positions of
vortices and regular pinning sites, respectively. We also assume that the
number of vortices is exactly equal to the number of pinning sites, i.e.,
filling factor is one.

\begin{figure}[btp]
\begin{center}
%\hspace*{-1.0cm}
%\hspace*{-0.5cm} 
%\includegraphics*[width=8.0cm]{1a.eps} %[bb=94 490 434 704, width=8.0cm, clip]
%\includegraphics*[width=15.0cm]{depf01.eps} %[bb=94 490 434 704, width=8.0cm, clip]
%\includegraphics*[width=6.5cm]{depf01cm.eps} %[bb=94 490 434 704, width=8.0cm, clip]
%%\includegraphics*[width=5.0cm]{depf01cm.eps} %[bb=94 490 434 704, width=8.0cm, clip]
\includegraphics*[width=8.5cm]{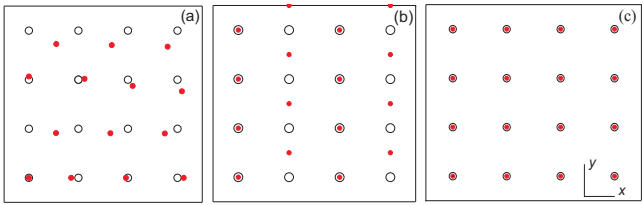} %[bb=94 490 434 704, width=8.0cm, clip]
\end{center}
\vspace{-0.5cm}
\caption{(Color online) Vortex lattice structure in a superconducting film
with weak periodic pinning at filling factor 1. Vortices are shown by red 
(dark gray) filled circles and pinning sites by black open cirles. 
Fig. 1(a) corresponds to the case of very low pinning with the deformed 
triangular vortex lattice. 
Fig. 1(b) shows a half-pinned regular phase realized at intermediate 
values of pinning strength. Fig. 1(c) represents a pinned vortex lattice, 
which is energetically favorable at higher values of pinning strength. }
\label{fig1}
\end{figure}

Let us recall first the situation when there is no random potential in the
system. In this case, as was explained in the Introduction, strong regular
pinning favors a square vortex lattice symmetry, see Fig. 1(c). At low $%
U_{0} $ the repulsion between vortices dominates resulting in a slightly
deformed triangular lattice that has the lowest energy, as illustrated in
Fig. 1(a). Finally, in the intermediate range of $U_{0}$, a half-pinned
lattice becomes the ground state, see Fig. 1(b). In the half-pinned phase,
an effective pinning potential for depinned vortices is created by their
interaction with neighboring rows of pinned vortices. It is easy to see that
the following expressions define the boundaries between the triangular
lattice and the half-pinned phase, and half-pinned phase and square lattice,
respectively \cite{Ukr,Pogosov}%
\begin{equation}
U_{0}\simeq 2\left( E_{hp}-E_{tr}\right)  \tag{6}
\end{equation}%
\begin{equation}
U_{0}=2\left( E_{sq}-E_{hp}\right) ,  \tag{7}
\end{equation}%
where $E_{tr}$, $E_{hp}$, and $E_{sq}$ are the energies of triangular,
half-pinned and square vortex lattices, without any pinning, taken per one
vortex. In other words, these vortex-vortex interaction energies depend only
on vortex lattice symmetry. We focus on the situation when the strength of a
regular pinning is not too high, i.e., is close to the boundaries defined by
Eqs. (6) and (7). In this case the vortex lattice can be considered as a
collective object and we are far from the regime of single-vortex pinning.

\subsection{Molecular-dynamics simulations}

In addition to an analytical approach, we supplement our study with computer
experiments. In this Subsection, we describe briefly the basic ingredients
of the used computer simulations. We use molecular-dynamics simulations of
vortices moving under the action of the forces due to the vortex-vortex
interaction and the interaction of vortices with regular and random pinning
sites. To find the lowest-energy vortex configurations, we perform simulated
annealing simulations by numerically integrating the overdamped equations of
motion %(see, e.g., Refs.~\cite{md01,md0157,md02,penrose}):
(see, e.g., Refs.~\cite{12,13,14,15,18}): 
\begin{equation}
\eta \mathrm{\mathbf{v}}_{i}\ =\ \mathrm{\mathbf{f}}_{i}\ =\ \mathrm{\mathbf{%
f}}_{i}^{vv}+\mathrm{\mathbf{f}}_{i}^{vp}+\mathrm{\mathbf{f}}_{i}^{T}. 
\tag{8}  \label{eqmd}
\end{equation}%
Here, $\mathrm{\mathbf{f}}_{i}$ is the total force per unit length acting on
vortex $i$, $\mathrm{\mathbf{f}}_{i}^{vv}$ and $\mathrm{\mathbf{f}}_{i}^{vp}$
are the forces due to the vortex-vortex and vortex-pin interactions,
respectively, and $\mathrm{\mathbf{f}}_{i}^{T}$ is the thermal stochastic
force. In Eq.~(8), %In Eq.~(\ref{1}), 
$\eta $ is the viscosity, which is set here to unity. The force due to the
interaction of the $i$-th vortex with other vortices is 
\begin{equation}
\mathrm{\mathbf{f}}_{i}^{vv}\ =\ \frac{2\pi }{\kappa ^{2}}\sum\limits_{j\neq
i}^{N_{v}}\ K_{1}\!\left( \mid \mathrm{\mathbf{r}}_{i}-\mathrm{\mathbf{r}}%
_{j}\mid \right) \hat{\mathrm{\mathbf{r}}}_{ij}\;,  \tag{9}  \label{fvv}
\end{equation}%
where $N_{v}$ is the number of vortices, $K_{1}$ is the modified Bessel
function, and $\hat{\mathrm{\mathbf{r}}}_{ij}=(\mathrm{\mathbf{r}}_{i}-%
\mathrm{\mathbf{r}}_{j})/\mid \mathrm{\mathbf{r}}_{i}-\mathrm{\mathbf{r}}%
_{j}\mid $. The modified Bessel function $K_{1}(r)$ decays exponentially for 
$r$ larger than 1 ($\lambda $ in dimensional units), thus it is safe to cut
off the (negligible) force for distances larger than $5$. In our
calculations, the logarithmic divergence of the vortex-vortex interaction
forces for $r\rightarrow 0$ is eliminated by using a cutoff at distances
smaller than $0.1$.

The force due to the interaction of the $i$th vortex with the parabolic
pinning potentials %(Eq.~(\ref{4})) 
(Eq.~(\ref{pinpot})) is: 
\begin{equation}
\mathrm{\mathbf{f}}_{i}^{vp}=\sum\limits_{k}^{N_{p}}\left( \frac{f_{p}}{%
\sigma }\right) \mid \mathrm{\mathbf{r}}_{i}-\mathrm{\mathbf{r}}%
_{k}^{(p)}\mid \Theta \!\left( r_{p}-\mid \mathrm{\mathbf{r}}_{i}-\mathrm{%
\mathbf{r}}_{k}^{(p)}\mid \right) \hat{\mathrm{\mathbf{r}}}_{ik}^{(p)}, 
\tag{10}  \label{fvp}
\end{equation}%
where $N_{p}$ is the number of pinning sites, $f_{p}$ is the maximal pinning
force of each potential well, $\sigma $ is the spatial range of the pinning
potential (for random potentials, we define $f_{pr}\lesssim f_{p}$ and $%
\sigma _{r}\lesssim \sigma $ as the maximum pinning force and radius,
correspondingly), $\Theta $ is the Heaviside step function, and $\hat{%
\mathrm{\mathbf{r}}}_{ik}^{(p)}=(\mathrm{\mathbf{r}}_{i}-\mathrm{\mathbf{r}}%
_{k}^{(p)})/\mid \mathrm{\mathbf{r}}_{i}-\mathrm{\mathbf{r}}_{k}^{(p)}\mid .$

The temperature contribution to Eq.~(\ref{eqmd}) is modeled by a stochastic
term obeying the following conditions: 
\begin{equation}
\langle f_{i}^{T}(t)\rangle =0,  \tag{11}
\end{equation}%
and 
\begin{equation}
\langle f_{i}^{T}(t)f_{j}^{T}(t^{\prime })\rangle =2\,\eta
\,k_{B}\,T\,\delta _{ij}\,\delta (t-t^{\prime }).  \tag{12}
\end{equation}%
The ground state of a system of vortices is obtained as follows. First we
set a high temperature to let vortices move randomly, then temperature is
gradually decreased down to $T=0$, thus simulating field-cooled experiments
(see, e.g. Refs.~\cite{harada,togawa}). Note that unlike in Refs.~\cite%
{12,13,14,15,18} we do not consider vortex dynamics driven by an external
Lorentz force, but we only calculate the ground-energy vortex configuration
(see, e.g., \cite{irina07,triangle}).

In Sec.~III we discuss the effect of a random potential on the
square and the half-pinned phases and define typical vortex lattice defects
induced by the randomness.

\section{Kinks}

Both square and half-pinned phases are periodic in the absence of random
pinning sites. Translational motion of the vortex lattice as a whole by a
period of a regular pinning lattice $a$ does not change the energy of the
system. Therefore, one can expect that there exist such solutions when, in
one region of the system, the vortex lattice is shifted by $a$ with respect
to that in another region. There should also be a kink with intermediate
value(s) for deviations of vortex positions separating these domains. In
this sense, our system resembles a traditional sine-Gordon system, which
describes the behavior of elastic media in a periodic potential \cite{Lub}.
For sine-Gordon systems, a typical kink has a smooth structure and its
length depends both on the elastic properties of the media and the strength
of the periodic potential, and it can be arbitrary large. Kinks are known to
play an important role in the process of disordering of media \cite{Lub}.
Thus one may say that disordering is occurring through a \textit{%
proliferation of kinks}. Therefore, we will pay special attention to the
kink structure.

\subsection{Continuous kinks versus sharp defects}

One can naively assume that there also exist smooth kinks for square pinned
and half-pinned lattices. However, despite the similarity with sine-Gordon
systems, our system has one specific feature, namely, square and half-pinned
vortex configurations are \textit{not} the lowest energy states in the
absence of pinning. Moreover, they are \textit{locally} unstable, i.e., even
infinitesimally small displacements of vortices lead to the collapse of the
lattice into a triangular array. In other words, the elasticity theory does
not work properly. This is shown explicitly in the Appendix, where we also
determine a structure of an infinite-lenght kink. We found that it has a
\textquotedblleft discrete\textquotedblright\ (i.e., not smooth) structure,
which is shown in Fig. 2(a). This is a fundamental difference between the
system at hand and conventional sine-Gordon systems.

Note that in the infinite system this type of defects has a divergent energy
because it is of infinitely long. Therefore, fluctuations generate \textit{%
reconnected} kinks of finite sizes rather than infinitely long defects, as
shown in Fig. 2 (a). This point will be explained in detail in Section V.
Also notice that the fact of appearance of reconnected defects is well-known
in the condensed matter physics, see e.g. Ref. \cite{Lub} for sine-Gordon
and related systems.

\begin{figure}[btp]
\begin{center}
%\hspace*{-1.0cm}
%\hspace*{-0.5cm} 
%\includegraphics*[width=6.0cm]{2a.eps} %[bb=94 490 434 704, width=8.0cm, clip]
%\includegraphics*[width=15.0cm]{depf02.eps} %[bb=94 490 434 704, width=8.0cm, clip]
%%\includegraphics*[width=7.5cm]{depf02cm.eps} %[bb=94 490 434 704, width=8.0cm, clip]
\includegraphics*[width=8.0cm]{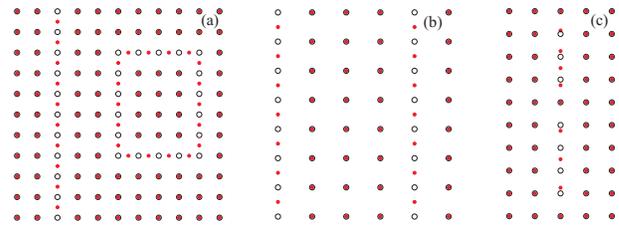} %[bb=94 490 434 704, width=8.0cm, clip]
\end{center}
\vspace{-0.5cm}
\caption{(Color online) Schematic view of typical defects in a regular
vortex lattice. 
Vortices/pinning sites are shown by red (dark gray) filled circles/black 
open cirles. 
Fig. 2(a) shows elastic strings of unpinned vortices in the
square pinned lattice: straight (on the left) and reconnected (on the right). 
Fig. 2(b) presents a row of pinned vortices in the half-pinned phase. 
Fig. 2(c) corresponds to the quasi-one-dimensional defect of finite
length inside the pinned square lattice.}
\label{fig2}
\end{figure}

It is important to notice that the string of unpinned vortices behaves as an
elastic object, vortices inside the string are correlated and behave
collectively due to their interaction. It is easy to see from Fig. 2(a) that
such a defect can be considered as a nucleus of a half-pinned vortex phase.
The defect energy density can be therefore estimated as a difference in the
energy density between the half-pinned and pinned vortex phases. They are
given, respectively, by%
\begin{equation}
\varepsilon _{hp}=\frac{E_{hp}-U_{0}/2}{a^{2}},  \tag{13}
\end{equation}%
\begin{equation}
\varepsilon _{sq}=\frac{E_{sq}-U_{0}}{a^{2}}.  \tag{14}
\end{equation}%
The first term in the numerator in the right-hand side (RHS) 
of Eq. (13) (Eq. (14)) is the
energy of the vortex-vortex interaction, taken per one vortex, while the
second term denotes the pinning energy per one vortex. The latter terms are
different for the half-pinned (Eq. (13)) and for pinned (Eq. (14)) phases,
since in the second case all the vortices are pinned, while in the first
case -- only half. Denominators in the RHS of Eqs. (13) and (14) give the
area per one vortex. Finally, for the defect energy density we obtain 
\begin{equation}
E_{d}^{(1)}\approx \varepsilon _{hp}-\varepsilon _{sq}=\frac{U_{0}/2-\left(
E_{sq}-E_{hp}\right) }{a^{2}}.  \tag{15}
\end{equation}%
It is important to note that the kink energy vanishes at the transition
between the square and half-pinned lattices, i.e., at $U_{0}=2\left(
E_{sq}-E_{hp}\right) $ since energies of these two phases become equal.

It is easy to realize that the pinned phase can also be disturbed by a
domain composed by a region of a deformed triangular vortex lattice. Again
such a domain is an elastic object, vortices within the domain are depinned
collectively. The region of depinned vortices has higher energy density
compared to the previously described string-like defect, since the energy of
a deformed triangular lattice has a higher energy than the half pinned
lattice in the relevant parameter region. This energy density can be
estimated in a similar manner, as the one which led us to Eq. (15). For that
we need to know the energy density of the deformed triangular lattice, which
is given by \cite{Pogosov} 
\begin{equation}
\varepsilon _{tr}\approx \frac{E_{tr}}{a^{2}},  \tag{16}
\end{equation}%
where $E_{tr}$ is the vortex-vortex interaction energy in the ideal
triangular lattice taken per one vortex. In reality, this lattice is
deformed and it is not ideal. However, as it was shown in Ref. \cite{Pogosov}%
, the energy of elastic distortions is compensated by the pinning energy in
the linear approximation with respect to vortices displacement, and Eq. (16)
is correct. Finally, for the density energy of the defect we obtain%
\begin{equation}
E_{d}^{(2)}\approx \varepsilon _{tr}-\varepsilon _{sq}=\frac{U_{0}-\left(
E_{sq}-E_{tr}\right) }{a^{2}}.  \tag{17}
\end{equation}

Now we turn to the description of the half-pinned phase. By similarity with
the pinned phase, we can see that a typical kink in the half pinned phase
consists of a row of pinned phase. One can also have domains of triangular
phases. The first type of defect is shown schematically in Fig. 2(b). The
energy densities for these kinks can be calculated from the differences in
energy densities of the nucleating phase and the initial one, as for Eqs.
(15) and (17). They are given, respectively, by 
\begin{equation}
E_{d}^{(3)}\approx \varepsilon _{sq}-\varepsilon _{hp}=\frac{-U_{0}/2+\left(
E_{sq}-E_{hp}\right) }{a^{2}},  \tag{18}
\end{equation}%
\begin{equation}
E_{d}^{(4)}\approx \varepsilon _{tr}-\varepsilon _{hp}=\frac{U_{0}/2-\left(
E_{hp}-E_{tr}\right) }{a^{2}}.  \tag{19}
\end{equation}%
which of the kinks has a lower-energy density depends on the position in
the phase diagram. As again can be expected, defects of the first type,
shown in Fig. 2(b), should be reconnected in finite systems.

\subsection{Quasi-one-dimensional defects}

We note that a one more type of the defects appears in the system for
sufficiently larger values of $U_{0}$, when the potential well for a pinned
vortex row is much deeper than the effective potential well created for a
depinned vortex row by surrounding rows of pinned vortices. In this limit,
the defect considered is a finite row of unpinned vortices, but \textit{not
reconnected}. Such a structure can exist due to the fact that in one part of
the defect, the string of vortices is stretched so that there is one missing
vortex and the last vortex in this part of the string deviates from its
equilibrium position by one period of the regular lattice $a$. In another
part of the defect, the string of vortices is compressed so that there is an
excess vortex. This defect is shown schematically in Fig. 2(c) for the most
simple case, when it is straight. It consists of a finite elastic string of
vortices in the same row, which are displaced along the direction of the
row. Thus this object as a whole can be considered as a \textit{%
quasi-one-dimensional kink-antikink structure}. In Section V, by using
molecular-dynamic simulations and rather general arguments, we argue that
both reconnected chains of depinned vortices and quasi-one-dimensional
defects are limiting cases of wider class of collective fractal-like
defects. In other words, defects considered in this Section are not
necessarily straight, but straight defects can be more easily studied
analytically.

We now calculate typical energy and length of this quasi-one-dimensional
defect. We assume that positions of all other vortices in the system remain
unchanged. The potential energy of a vortex in the string is given by the
interaction energy with the fixed vortices, with pinning sites, and also
with each other. The first contribution to the string energy can be found
from Eq. (3), where one can neglect all the terms except of the ones with $%
m=0$, $\pm 1$\ provided that $a\sim 1$, which results in%
\begin{equation}
E_{1}^{(1D)}\simeq \frac{4\pi }{\kappa ^{2}}\frac{\pi }{a}\sum_{k}\left( 
\frac{\exp \left( -a\right) }{1-\exp \left( -a\right) }+\frac{a}{\pi }\exp
\left( -2\pi \right) \cos \frac{2\pi y_{k}}{a}\right) ,  \tag{20}
\end{equation}%
where the summation is performed over positions of all vortices in the
string $y_{k}$. Note that from this equation one can see that the
intervortex interaction favors a shift of $a/2$\ of the row in the $y$
direction with respect to the lattice making a symmetry closer to the
triangular one. Note that the potential energy (20) for each vortex has a
typical sine-Gordon structure.

\begin{figure}[btp]
\begin{center}
%\hspace*{-1.0cm}
\hspace*{-0.5cm} 
\includegraphics*[width=8.0cm]{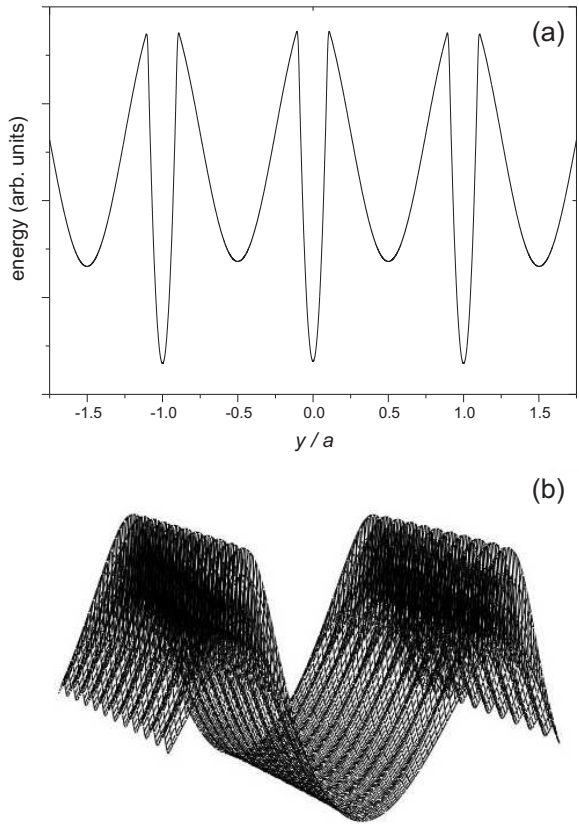} 
%[bb=94 490 434 704, width=8.0cm, clip]
\end{center}
\par
%\vspace{-1.5cm}
\vspace{-0.5cm} %\vspace{-0.5cm}
\caption{Potential energy profile for the vortex row in the
square lattice along one of the principal axis, provided that positions of
vortices in the row are fixed with respect to each other and all other
vortices are pinned. Fig 3(a) corresponds to the dependence on one
coordinate, whereas Fig. 3(b) shows a dependence on both coordinates, which
is of quasi-one-dimensional form.}
\label{fig3}
\end{figure}

The interaction energy for the vortices with pinning sites in the row reads 
\begin{equation}
E_{2}^{(1D)}=\sum_{n=-\infty }^{\infty }\sum_{k}V(an-y_{k}),  \tag{21}
\end{equation}%
where $V(y)$ is defined by Eq. (4) and the index $n$ stands for the
positions of the pinning sites. For illustrative purposes, we plot the sum
of $E_{1}^{(1D)}$ and $E_{2}^{(1D)}$ in Fig. 3(a) for vortices in the string
that are positioned periodically, $y_{k}=y+ka$. The resulting potential is,
of course, a periodic function in $y$. The energy minimum for the string is
attained for a square configuration due to our choice of parameters, i.e.,
for strong enough regular pinning. To show that this potential is of a
quasi-one-dimensional form, we also plot it in three dimensional form in
Fig. 3(b).

There is an additional contribution to the energy of the vortex, which is
due to its interaction with other vortices in the string, i.e., the
elasticity term. This contribution depends on the mutual positions of the
vortices. We will take into account only the interaction with nearest
neighbors, as is usually done in elasticity theory. Taking into account Eq.
(2) and using a Taylor expansion for the modified Bessel function, we obtain
the following expression for the energy of the whole string of vortices

\begin{equation}
E_{3}^{(1D)}=\frac{2\pi }{\kappa ^{2}}\left( K_{0}\left( a\right) +\frac{%
K_{1}\left( a\right) }{a}\right) \sum_{k}\left( \Delta u_{k}\right) ^{2}, 
\tag{22}
\end{equation}%
where $\Delta u_{k}$\ is the displacement of $k$th vortex with respect to
the $(k-1)$th one. As usual, we can replace $\Delta u_{k}/a$\ by $du/dy$ and
also switch from the summation to the integration in the RHS of Eq. (22) 
\cite{Lub}:

\begin{equation}
E_{3}^{(1D)}=\frac{2\pi a}{\kappa ^{2}}\left( K_{0}\left( a\right) +\frac{%
K_{1}\left( a\right) }{a}\right) \int \left( \frac{du}{dy}\right) ^{2}dy, 
\tag{23}
\end{equation}%
where the integration is performed along the length of the defect. The total
energy of the defect is thus given by the sum of three contributions:%
\begin{equation}
E^{(1D)}=E_{1}^{(1D)}+E_{2}^{(1D)}+E_{3}^{(1D)},  \tag{24}
\end{equation}%
which can be estimated in a simple manner. We assume that a typical
kink-antikink structure has a length $l_{1D}$. There are $l_{1D}/a$ vortices
along this defect and they are depinned from their lowest energy positions.
Each vortex gains some energy by being depinned. If the width of the
potential well $\sigma $ is much smaller than the intersite distance $a$,
the gain is given by $U_{0}$, as follows from Eq. (23). At the same time,
each vortex also loses some amount of energy due to the fact that it no
longer forms a square configuration with vortices from the surrounding fixed
rows, as can be seen from Eq. (20). Thus, the total energy increase for the
string of $l_{1D}/a$ depinned vortices is given approximately by%
\begin{equation}
E_{1}^{(1D)}+E_{2}^{(1D)}\approx l_{1D}/a\left( U_{0}-\frac{4\pi }{\kappa
^{2}}e^{-2\pi }\right) .  \tag{25}
\end{equation}%
The elastic energy of the string $E_{3}^{(1D)}$ can be easily estimated from
Eq. (23) as%
\begin{equation}
E_{3}^{(1D)}\approx \frac{2\pi a^{3}}{\kappa ^{2}}\frac{1}{l_{1D}}\left(
K_{0}\left( a\right) +\frac{K_{1}\left( a\right) }{a}\right) .  \tag{26}
\end{equation}

If we now minimize the total defect's energy $E^{(1D)}$ with respect to its
length $l_{1D}$, we obtain%
\begin{equation}
l_{1D}\approx \sqrt{\frac{2\pi a^{4}}{\left( U_{0}-\frac{4\pi }{\kappa ^{2}}%
e^{-2\pi }\right) \kappa ^{2}}\left( K_{0}\left( a\right) +\frac{K_{1}\left(
a\right) }{a}\right) }.  \tag{27}
\end{equation}%
This leads to the following result for the defect energy%
\begin{equation}
E^{(1D)}\approx 2\sqrt{\frac{2\pi a^{2}\left( U_{0}-\frac{4\pi }{\kappa ^{2}}%
e^{-2\pi }\right) }{\kappa ^{2}}\left( K_{0}\left( a\right) +\frac{%
K_{1}\left( a\right) }{a}\right) },  \tag{28}
\end{equation}%
which has a nontrivial dependence on $U_{0}$. As we will see below, in the
limit of weak pinning, quasi-1D defects have higher energies compared to
previously analyzed ones. This can be explained by noticing that vortices in
the string are situated neither exactly in a potential well induced by
pinning sites nor in that created by the neighboring rows of vortices.

Note that a quasi-1D defect can have a slightly different structure, when
the maximum deviation of vortex in the string from its equilibrium position
is given by $a/2$ but not by $a$. In this case, the position of the most
strongly displaced vortex in the row is stabilized by the interaction with
neighboring rows of vortices and not by a pinning site. Our estimates for
length and energy of quasi-one-dimensional defect, given by Eqs. (27) and
(28), however remain applicable also for such a defect.

\section{Random potential}

In this Section, we characterize the random potential and describe how to
deal with it in our problem. We assume that the random potential is created
by randomly distributed pinning wells with size $\sigma _{r}\lesssim $ $%
\sigma $, $a$ and depth $U_{r}\lesssim $ $U_{0}$. The concentration of
random sites is denoted by $n_{r}$, and we also assume that it is higher or
equal to the concentration of regular pinning sites, $n_{r}\gtrsim a^{-2}$.

It is well-known that the actual pinning energy due to the random pinning is
produced by fluctuations of the random potential in a given region of the
system, i.e. by bunches of pins, and not by pinning potential averaged
value. If we have a region of area $S_{0}\gg n_{r}^{-1}$, the distribution
probability $P(N)$ for having $N$ random pinning sites within this region
can be well described by a normal distribution%
\begin{equation}
P(N)=\frac{1}{\sqrt{2\pi }}\frac{1}{\sqrt{n_{r}S_{0}}}\exp \left[ -\frac{%
(N-n_{r}S_{0})^{2}}{2n_{r}S_{0}}\right] .  \tag{29}
\end{equation}%
As can be expected, the averaged value of $N$ is given by%
\begin{equation}
\left\langle N\right\rangle =n_{r}S_{0},  \tag{30}
\end{equation}%
whereas the mean-square deviation reads%
\begin{equation}
\sqrt{\left\langle N^{2}\right\rangle -\left\langle N\right\rangle ^{2}}%
\simeq n_{r}S_{0}.  \tag{31}
\end{equation}%
The probability to have any number of random pinning sites larger than some
definite value $N_{0}$ within the region $S_{0}$ can be found by
integration, $\int_{N_{0}}^{\infty }P(N)dN$, and this integral can be
estimated as (provided that $N_{0}-n_{r}S_{0}\gtrsim \sqrt{n_{r}S_{0}}$)%
\begin{equation}
\int_{N_{0}}^{\infty }P(N)dN\approx \frac{1}{\sqrt{2\pi }}\frac{\sqrt{%
n_{r}S_{0}}}{N_{0}-n_{r}S_{0}}\exp \left[ -\frac{(N_{0}-n_{r}S_{0})^{2}}{%
2n_{r}S_{0}}\right] .  \tag{32}
\end{equation}%
The probability for this event not to occur is, of course, $%
1-\int_{N_{0}}^{\infty }P(N)dN$. If we now have a larger region $S>S_{0}$,
the probability that, within $S$, one cannot find a region of area $S_{0}$\
with number of random pinning sites larger than $N_{0}$ is estimated as $%
\sim \left( 1-\int_{N_{0}}^{\infty }P(N)dN\right) ^{S/S_{0}}$. Finally, the
probability to find such a region is $1-\left( 1-\int_{N_{0}}^{\infty
}P(N)dN\right) ^{S/S_{0}}$. From this analysis and from Eq. (32) it is easy
to see that if $S\sim S_{0}$ and $\left\vert N_{0}-n_{r}S_{0}\right\vert
\sim \sqrt{n_{r}S_{0}}$, this probability is very close to 1. This means
that within the domain of area $S$ slightly larger than $S_{0}$ we can
always find a bunch of pines, such that the pinning energy around this bunch
(within region of area $S_{0}$) deviates by $\sqrt{n_{r}S_{0}}$ from its
average value. This illustrates that the effective pinning energy within
some region can be closely approximated by mean-square deviation of random
pinning potential within this region, $U_{r}\sqrt{n_{r}S_{0}}$. This
conclusion is well-known in the physics of disordered flux-line lattices 
\cite{Larkin}. The famous Larkin-Ovchinnikov theory is also based on the
idea that the mean-square deviation of random pinning energy is balanced
with the energy of lattice distortions within some domain.

Each kind of defects described in Section IV has its own typical energy and
sizes. In order to estimate the random pinning strength leading to a
proliferation of such defects one can equate the defect energy to the
mean-square deviation of the random pinning potential within the
characteristic area of the defect. By doing this, one can study the phase
diagram of the system in the plane of random and regular pinning sites
strengths.

\section{Phase diagram}

In the absence of random pinning, there are two phase boundaries separating
the pinned square, half pinned and deformed triangular lattices, as given by
Eqs. (6) and (7). We now consider the effect of the random pinning
potential. Note that a deformed triangular vortex lattice is aperiodic even
in the absence of random pinning, since periods of vortex and regular
pinning sites lattices are incommensurate. Therefore we consider disordering
of square pinned and half-pinned vortex lattices, which have a regular
structure. As was already explained, the destruction of the order occurs via
a proliferation of defects. Using the tools developed in Secs.~III and IV, 
we can estimate the typical random pinning strengths leading to a
generation of different defects, identified in Section III. We do not expect
that this will give us the full phase diagram, since we have found defects
only in some simple and limiting cases. In order to obtain a deeper insight
to the problem we will also use molecular dynamic simulations.

Let's start from the case of a square pinned lattice. The lowest energy
defect in such a situation, among those we have found, is constructed from
the elastic string of collectively depinned vortices, as explained in
Section III. This row has to be reconnected because otherwise its energy is
infinitely large \cite{Lub} (therefore, one should pay a special attention
to this point both in numerical and real experiments on finite arrays, since
there it is easy to get just straight defects - their energies remain finite
because the system sizes are finite). The radius of such a defect cannot be
too small, since the system is discrete. Therefore, the lowest energy defect
has a radius $R_{d}$ of the order of the intersite distance, $R_{d}\sim a$,
see Fig. 2 (a). Let us write $R_{d}$ as $ar_{d}$, where the dimensionless
factor $r_{d}$ is of the order of one, $r_{d}\sim 1$. The energy of such a
defect can be found from Eq. (15):%
\begin{equation}
E_{kink}\sim 2\pi r_{d}\left[ U_{0}/2-\left( E_{sq}-E_{hp}\right) \right] . 
\tag{33}
\end{equation}%
The area of this structure is just $\pi a^{2}r_{d}^{2}$. To find the random
pinning strength leading to a proliferation of this particular kind of
defects, we equate the energy of the defect $E_{kink}$ to the mean-square
deviation of the random pinning energy within the area of the defect\ $U_{r}%
\sqrt{n_{r}\pi a^{2}r_{d}^{2}}$ and we find%
\begin{equation}
U_{r}\sim \sqrt{\frac{\pi }{n_{r}a^{2}}}\left[ U_{0}-2\left(
E_{sq}-E_{hp}\right) \right] .  \tag{34}
\end{equation}%
From Eq. (34) we see that the $U_{r}$ required to destroy the order in the
vortex positions in the square pinned phase vanishes at $U_{0}=2\left(
E_{sq}-E_{hp}\right) $, i.e., in the point of the transition between the
square pinned and half-pinned lattices in the absence of disorder. This
means that arbitrary weak disorder destroys the regularity in a square
vortex lattice in the vicinity of this point, since the energies of pinned
and half pinned lattices are equal to each other there. 

We now start to construct a phase diagram of the system in the $U_{0}$ - $%
U_{r}$ plane for the same concentration of random and regular pins, i.e., $%
n_{r}a^{2}=1$. The boundary separating pinned and a mixture of pinned and
half-pinned lattices is shown by line 1 in Fig. 4 at $a=1$ and $\kappa =5$.
This mixed phase still can be considered as pinned, as a whole, by the
regular pinning sites array, since there is a strong correlation between the
positions of regular sites and the vortices.

\begin{figure}[btp]
\begin{center}
\vspace{-0.5cm} 
%\hspace*{-1.0cm}
\hspace*{-0.5cm} 
\includegraphics*[width=10.0cm]{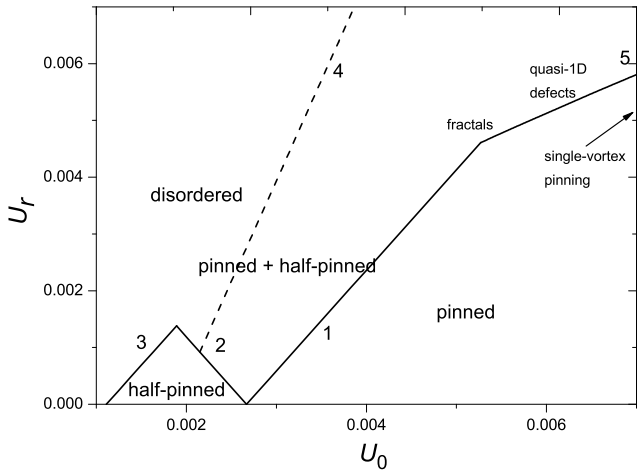} %[bb=94 490 434 704, width=8.0cm, clip]
\end{center}
%\vspace{-0.5cm} 
\vspace{-1.0cm} 
%\vspace{-1.5cm} 
\caption{ Schematic phase diagram of the vortex lattice in the plane of the
periodic pinning strength $U_{0}$ and the random pinning strength $U_{r}$. }
\label{fig4}
\end{figure}

Let us now switch to the half-pinned phase. As was explained in Section III,
the lowest energy defect in this case corresponds to the row of pinned
vortices, which recovers locally square array, as in Fig. 2(b). This row
again has to be reconnected. By repeating the derivation which led us to Eq.
(34), we obtain the following criterion for $U_{r}$ favoring proliferation
of such defects%
\begin{equation}
U_{r}\sim \sqrt{\frac{\pi }{n_{r}a^{2}}}\left[ -U_{0}+2\left(
E_{sq}-E_{hp}\right) \right] .  \tag{35}
\end{equation}%
It corresponds to the curve 2 in the phase diagram which is basically a
mirror image of the boundary given by Eq. (34). It is interesting to note
that \textit{disorder} in this case leads to an unexpected situation,
namely, it forces vortices within some regions to be pinned by the \textit{%
regular }array of pinning sites. Also note that if we follow along curves 1
and 2, we see that the minimal random pinning strength required for the
partial disordering of the vortex lattice is a non-monotonic function of $%
U_{0}$. This is again a nontrivial result, since one could expect that the
higher $U_{0}$, the higher must be $U_{r}$, which destroys the order in the
vortex positions. However, this is not exactly the case, since there is one
more factor, i.e., the intervortex interaction, which helps the random
pinning sites to destroy order. Thus, at a certain value of $U_{0}$ it is
energetically favorable for the system to decrease its energy not by
introducing vortex lattice defects, but by making a structural phase
transition of the vortex lattice, on large scales, into a half-pinned 
\textit{regular} phase.

The half-pinned phase can also be destroyed via elastic domains of the
deformed triangular lattice; the energy density for the corresponding defect
is given by Eq. (19). Thus, the random energy $U_{r}$ generating such
domains can be estimated as%
\begin{equation}
U_{r}\sim \sqrt{\frac{\pi }{n_{r}a^{2}}}\left[ U_{0}-2\left(
E_{hp}-E_{tr}\right) \right] .  \tag{36}
\end{equation}%
It is shown as curve 3 in Fig. 4, which separates regions of half-pinned and
deformed triangular, i.e., completely disordered, phase. Let us stress that
proliferation of such domains leads to a complete depinning of the vortex
lattice from the underlying square array of pinning sites. Curves 2 and 3
intersect in some point and they bound a region in the phase diagram where
the half pinned vortex lattice is stabilized. Of course, this does not mean
that there are no defects at all in this region (or in the region of the
square pinned lattice), but this implies that their concentration is low.
One should also understand that the state, which we call a mixture of pinned
and half pinned phases, contains domains with the deformed triangular
lattice as well, but again the concentration of such domains is low. There
can be other types of defects, like vacancies, when one vortex is
redistributed to the position which is very far, in terms of lattice period,
from its initial position, however, the probability of such events is low,
since we are in the regime of collective pinning. Therefore, such events do
not contribute significantly to the process of vortex lattice disordering.

Now we discuss how a mixture of pinned and half pinned phases is finally
completely depinned from the regular pinning array. For that we first
estimate $U_{r}$ generating domains of triangular lattice inside the \textit{%
square} one: 
\begin{equation}
U_{r}\sim 2\sqrt{\frac{\pi }{n_{r}a^{2}}}\left[ U_{0}-\left(
E_{sq}-E_{tr}\right) \right] ,  \tag{37}
\end{equation}%
which is depicted as curve 4 in Fig. 4. In reality, before such domains
appear in the square phase, it is already disturbed by rows of the
half-pinned lattice, so that curve 4 does not describe properly such a
transition. The same is true for the curve 3 which describes the transition
from the half pinned to the deformed triangular lattice. In other words,
curves 3 and 4 must intersect. 

From the obtained phase diagram we see that the disordering of the square
pinned lattice in the limit of weak pinning always occurs in two steps.
First, some rows of vortices depin and form reconnected elastic chains;
their concentration increases with the increase of random pinning strength $%
U_{r}$. At higher values of $U_{r}$, elastic domains of deformed triangular
lattice proliferate in the system leading to a complete disordering of
vortex array. If we start from the half pinned lattice, two scenarios are
possible. If the regular pinning strength $U_{0}$ is relatively low, then
the order is destroyed in one step due to the proliferation of domains of
the deformed triangular lattice. If $U_{0}$ is relatively high, again a
two-step disordering can take place. As a first step, an admixture of the
square pinned lattice appear in the system, and as a second step, domains of
a deformed triangular lattice depin the vortex lattice from the regular
pinning array. Note that in experimental work \cite{French} on macroscopic
charged balls on a square array of traps, the formation of domains in the
lattice of these balls was also observed. Although there was no artificial
randomness in the experimental setup, the disorder was certainly
unavoidable. Since energies of three relevant vortex phases, namely,
triangular, pinned, and half-pinned ones, are quite close to each other,
domains formation can be observed in such experiments.

In Section III, we also have described a quasi-1D type of defects, which
destroy the order in vortex positions within the same row, see Fig. 2(c).
The typical area of such a defect is $\sim l_{1D}a$, where $l_{1D}$ is given
by Eq. (27). The energy of the defect is defined in Eq. (28). Thus, for $%
U_{r}$ \ leading to the generation of quasi-1D defects we get 
\begin{equation}
U_{r}\sim \frac{2}{\sqrt{a}}\frac{2\pi }{\kappa ^{2}}\left( U_{0}-\frac{4\pi 
}{\kappa ^{2}}e^{-2\pi }\right) ^{3/4}\left[ K_{0}\left( a\right) +\frac{%
K_{1}\left( a\right) }{a}\right] ^{1/4}.  \tag{38}
\end{equation}%
Our estimates show that, for realistic values of the parameters, $U_{r}$
could become larger than the values corresponding to the phase boundary
depicted by lines 2 and 3. This conclusion is due to the fact that quasi-1D
defects represent an intermediate case between the collective pinning of
vortices and a regime of single-vortex pinning at high pinning strengths.
The vortex lattice in this case is still an elastic media, with the
intervortex interaction being important in the total balance of the
energies. The phase boundary corresponding to such a transition is plotted
in Fig. 4 as line 5. Since quasi-1D defects destroy the order in the vortex
positions in the same row, a mixture of pinned and half-pinned phases in
this case is also characterized by the absence of such an order in the rows
of unpinned vortices; moreover, chains of unpinned vortices are now not
necessarily reconnected. In this sense, the difference between the quasi-1D
disordered phase and a mixture of pinned and half-pinned states is smeared
out.

One can expect that the final regime with sufficiently high $U_{0}$
corresponds to zero-dimensional defects, when vortices are depinned
individually that is known as the single-vortex pinning regime. Thus, the
dimensionality of typical defects leading to the destruction of the order is
decreasing continuously from 2 to 0 when the regular pinning strength $U_{0}$
is increasing. At low $U_{0}$, regularity in the vortex positions is
destroyed by the appearance of reconnected elastic chains of
collectively-depinned vortices that are ordered within each chain. At higher 
$U_{0}$, these chains start to be open, they should become straighter and, 
therefore, one can expect neighboring chains to connect due to
the lack of free space. If we again increase $U_{0}$, chains become
quasi-one dimensional elastic rows of finite length, as we have shown
schematically in Fig. 2(c). Finally, we are in a single-pinning regime, when
each vortex is pinned individually, that can be interpreted as chain
shrinking towards a single lattice period, since the chain length given by
Eq. (27) decreases with the increase of regular pinning strength.

So far only limiting cases of the first stage of vortex lattice disordering
have been considered, namely, when we have: (i) reconnected chains, i.e.,
two-dimensional collective defects, (ii) quasi-one-dimensional stripe-like
collective defects, (iii) quasi-zero-dimensional individual defects in the
regime of a single-vortex pinning. To have a deeper insight into the problem
and to understand better how these defects evolve, we perform
molecular-dynamics simulations for the regime intermediate between the weak
and strong pinning regimes. In Figs. 5-8 we present typical vortex patterns
corresponding to an increasing strength of the random pinning potential $%
U_{r}$ with a constant concentration of random sites, which is set equal to
the concentration of regular pins. The initial state of the system (i.e.,
without random pinning) is a square lattice of vortices pinned on regular
pins. A simulation region with the sizes 20$\times $20 $\lambda $ contains
an array of 20$\times $20 regular pinning arrays and 400 vortices. We use
periodic boundary conditions at the boundaries of the simulation region. The
value of $U_{0}$ is $0.045\times 2\pi /\kappa ^{2}$, whereas the intervortex
distance is 1 in units of $\lambda $. This value of $U_{0}$ is several times
larger than that required for the transition from the square phase to the
half-pinned state, i.e., we are indeed in the intermediate regime, where
both reconnected and open chains of depinned vortices are expected to appear
in the system with increasing $U_{r}$. This is exactly what we see in Fig. 5
at $U_{r}=0.72U_{0}$, which contains both types of chains, and still there
is no domains of a deformed triangular lattice. In general, these elastic
chains are packed into finite fractal-like structures with a
quasi-self-similar topology and dimensionality between 1 and 2. The building
blocks of these fractals are represented by parts of rectangles of discrete
sizes (i.e., commensurate with the period of the pinning array). If we
increase $U_{r}$, as in Fig. 6 for $U_{r}=0.85U_{0}$, domains of deformed
triangular lattice start to appear in the system, and thus they coexist with
chains of unpinned vortices. However, these domains somehow suppress the
networks of depinned vortices by simply cutting them, since there is a
competition between these two kinds of defects for a free space in the
system. Further increasing $U_{r}$ leads to the increase of the
concentration of domains of totally depinned vortices, as indicated in Fig.
7 at $U_{r}=1.28U_{0}$. It is interesting to note that domains of pinned
square vortex lattice turn out to be rather robust with respect to the
random pinning; some of them survive up to quite large values of $U_{r}$,
several times greater than that corresponding to the situation shown in Fig.
7, with the fraction of pinned vortices decreasing slowly with increasing $%
U_{r}$. Fig. 8 shows the final state with quite high value of $U_{r}=5.3U_{0}
$, when almost all the vortices are depinned from the regular pinning array
and there is no correlation between the positions of the regular pins and
vortices, but there is a pronounced correlation between the positions of the
vortices and random pins. Note that typical values of $U_{r}$ leading to the
generation of defects are in good qualitative agreement with our analytical
predictions.

We also performed simulations for even higher $U_{0}$. We found that indeed
the initial defects turn out to be quasi-one-dimensional stripes with the
characteristic length of few lattice periods, rather than long-length and
branched structures. Other defects found correspond to higher values of $%
U_{r}$, and they are more similar to a single-particle imperfections than to
regions of elastically distorted media, which are constructed of many
interacting particles. In fact, these defects correspond to domains of
totally depinned vortices described above, in the limiting case of a
single-vortex pinning. Finally, single-vortex defects and stripe-like
defects also match, when the length of the stripes tends to a one lattice
period. A general qualitative picture of the evolution of defects is
presented in Fig. 9.

Thus, the molecular-dynamics simulations enable us to relate defects of
various dimensionalities and to better understand their evolution.

\begin{figure}[btp]
\begin{center}
\vspace{-0.5cm} 
%\hspace*{-1.0cm}
\hspace*{-0.5cm} 
\includegraphics*[width=9.0cm]{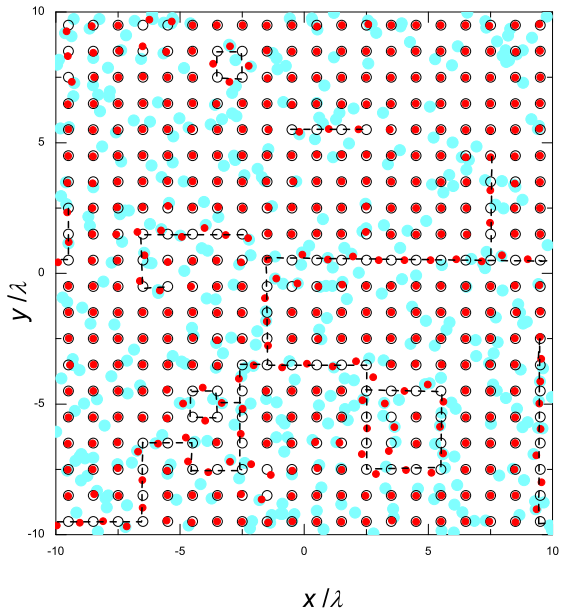} %[bb=94 490 434 704, width=8.0cm, clip]
\end{center}
%\vspace{-0.5cm} 
\vspace{-1.0cm} 
\caption{(Color online) Vortex pattern in the film with regular square and
random pinning arrays for $U_{r}=0.72U_{0}$, obtained in molecular-dynamics
simulations. Dashed lines are guides for eyes indicating positions of chains
of unpinned vortices. Irregular blue (light gray) spots represent positions 
of random pins, red (dark gray) filled circles show positions of vortices, 
and regular black open circles correspond to periodic pins.} 
\label{fig5}
\end{figure}

\begin{figure}[btp]
\begin{center}
\vspace{-0.5cm} 
%\hspace*{-1.0cm}
\hspace*{-0.5cm} 
\includegraphics*[width=9.0cm]{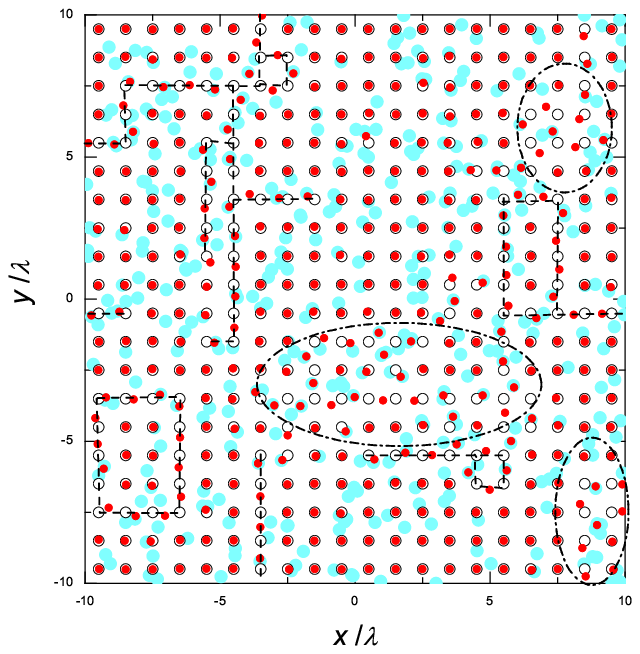} %[bb=94 490 434 704, width=8.0cm, clip]
\end{center}
%vspace{-0.5cm} 
\vspace{-1.0cm} 
\caption{(Color online) Vortex pattern in the film with regular square and
random pinning arrays for $U_{r}=0.85U_{0}$, obtained in molecular-dynamics
simulations. Dashed lines are guides for eyes indicating positions of chains
of unpinned vortices. Dash-dotted lines show borders of domains of deformed
triangular lattice which is depinned from the underlying regular pinning
array. Irregular blue (light gray) spots represent positions of random pins, 
red (dark gray) filled circles show positions of vortices, and regular black 
open circles correspond to periodic pins.} 
\label{fig6}
\end{figure}

\begin{figure}[btp]
\begin{center}
\vspace{-0.5cm} 
%\hspace*{-1.0cm}
\hspace*{-0.5cm} 
%\hspace*{-0.5cm} 
%\includegraphics*[width=12.0cm]{depf07.eps} %[bb=94 490 434 704, width=8.0cm, clip]
\includegraphics*[width=9.0cm]{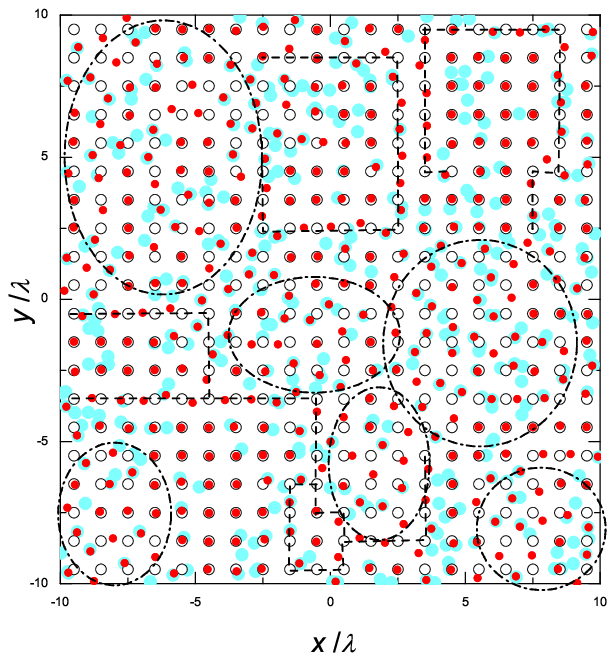} %[bb=94 490 434 704, width=8.0cm, clip]
\end{center}
%\vspace{-0.5cm} 
\vspace{-1.0cm} 
\caption{(Color online) Vortex pattern in the film with regular square and
random pinning arrays for $U_{r}=1.28U_{0}$, obtained in molecular-dynamics
simulations. Dashed lines are guides for eyes indicating positions of chains
of unpinned vortices. Dash-dotted lines show borders of domains of deformed
triangular lattice which is depinned from the underlying regular pinning
array. Irregular blue (light gray) spots represent positions of random pins, 
red (dark gray) filled circles show positions of vortices, and regular black 
open circles correspond to periodic pins.} 
\label{fig7}
\end{figure}

\begin{figure}[btp]
\begin{center}
\vspace{-0.5cm} 
%\hspace*{-1.0cm}
\hspace*{-0.5cm} 
\includegraphics*[width=9.0cm]{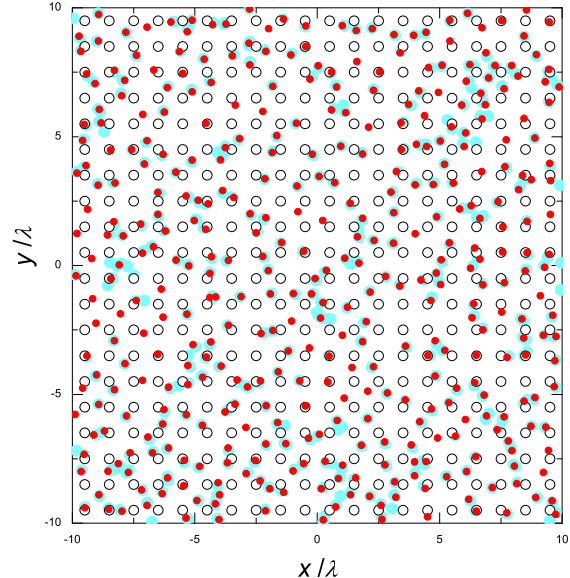} %[bb=94 490 434 704, width=8.0cm, clip]
\end{center}
%\vspace{-0.5cm} 
\vspace{-1.0cm} 
\caption{(Color online) Vortex pattern in the film with regular square and
random pinning arrays for $U_{r}=5.3U_{0}$, obtained in molecular-dynamics
simulations. Irregular blue (light gray) spots represent positions of random 
pins, red (dark gray) filled circles show positions of vortices, and regular 
black open circles correspond to periodic pins.} 
\label{fig8}
\end{figure}

%\begin{widetext}

\begin{figure}[btp]
\begin{center}
\vspace{0.5cm} 
%\hspace*{-1.0cm}
\hspace*{-0.5cm} 
\includegraphics*[width=8.0cm]{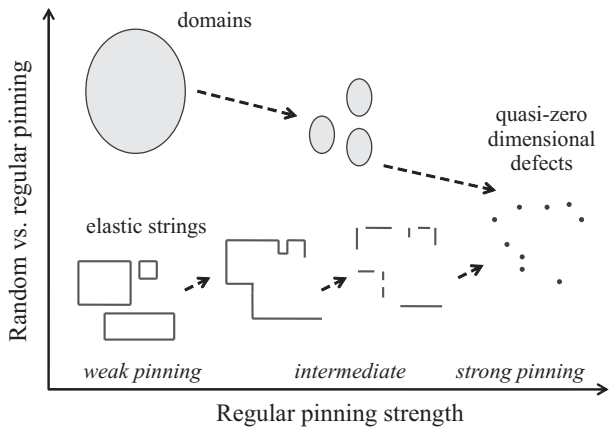} 
%[bb=94 490 434 704, width=8.0cm, clip]
\end{center}
\par
\vspace{-0.5cm} 
\caption{General picture of the evolution of disorder-induced
defects for square pinned vortex lattice. Lower branch corresponds to the
elastic strings of collectively depinned vortices. First, they appear as
reconnected chains. Then chains break and straigthen. As a result, they
start to overlap and form fractal-like clusters. Finally, they decay to
quasi-one-dimensional stripes, which shrink gradually to
quasi-zero-dimensional individual defects corresponding to the
single-pinning regime. Upper branch shows the evolution of higher-energy
elastic defects, which form domains of vortices collectively depinned from 
regular pins. These domains break into uncorrelated blocks of smaller sizes
and finally also shrink to quasi-zero dimensional defects thus matching with
the lower branch of defects.}
\label{fig9}
\end{figure}

%\end{widetext}

\section{Conclusions}

In this paper, we studied theoretically the effect of disorder on vortex
lattices in the presence of a regular square array of pinning sites. We
considered a two-dimensional system with the same concentration of regular
pinning sites and vortices, i.e., with filling factor 1. We were mostly
focused on a weak pinning regime, when vortex lattice demonstrates
collective properties due to the mutual repealing of vortices. The resulting
structure of the vortex lattice is determined by the competition between the
square symmetry of the regular pinning array, by the triangular symmetry
imposed by the vortex-vortex interaction, and by the action of randomness
trying to destroy the order in vortex positions. Typical defects of vortex
lattices have been identified for some limiting and therefore simple cases
and their energies and sizes have been estimated analytically.
Molecular-dynamic simulations were also used for exploring regions on phase
diagram, which are difficult to approach in an analytical way. Topological
defects found are of various dimensionalities including fractional ones
having fractal-like structure. We were able to predict analytically two
limiting cases of these fractal-like structures: a reconnected chain of
unpinned vortices and a straight stripe-like defects, while molecular
dynamic simulations allowed us to reveal the connection between these limits.

We then constructed an approximate phase diagram of the system in the plane
of random and regular pinning strengths. We found that disordering of the
regularity in vortex positions can occur either in one or two steps
depending on the strength of the regular pinning potential. In the former
case, domains of deformed triangular lattice start to appear spontaneously
in the film. In the latter case, in the beginning, reconnected chains of
unpinned vortices are generated inside a square lattice (or the opposite:
chains of pinned vortices appear inside the half-pinned phase), and only
after that again domains of deformed triangular lattice depin the vortex
lattice as a whole from the regular pinning array. Both chains and domains
are elastic defects with interacting vortices depinned collectively. 

We also analyze an intermediate regime for values of regular pinning
strength, which are between the regimes of collective and single-vortex
pinning. In this case, the first step of disordering occurs by proliferation
of opened, not reconnected, elastic chains of depinned vortices, which can
form long fractal-like defects with complicated quasi-self-similar topology.
Fractals appear due to the process of opening of reconnected chains and
their straightening, when different chains start to connect to each other.
At the same time, the average length of these defects is going to decrease
with the increase of regular pinning strength due to a periodic potential
sharpening, so that there is a competition between two tendencies. As a
result, chains become straighter and relatively short thus forming
quasi-one-dimensional stripes. Finally, for high values of periodic pinning
strength, they shrink continuously towards a quasi-zero-dimensional defects
corresponding to a single-pinning regime. The latter regime is characterized
by one-step disordering, i.e., the width of the region on the phase diagram
with coexistent chains of unpinned vortices and domains of totally depinned
vortices has to vanish with the growth of regular pinning strength, since
these two types of defects also match. Thus, we can understand on the same
footing all the regimes of disordering starting from a collective pinning to
a single-vortex pinning.

Our results are applicable not only to vortices in superconductors, but 
also for
a broader class of physical systems containing a large number of repealing
objects and underlying system of traps: for instance, vortices in ultracold
gases in optical lattices, colloids, macroscopic elastic balls interacting
via Coulomb forces, charge and spin density waves. 

\section{Acknowledgments}

This work was supported by the Flemish Science Foundation (FWO-Vl), the 
Interuniversity Attraction Poles (IAP) Programme -- Belgian State -- Belgian
Science Policy, and the ``Odysseus'' program of the Flemish Government and
the FWO-Vl. 
W.V.P. also acknowledges support from the Russian Foundation for Basic 
Research (project No.~06-02-16691), the President of Russia Programme 
for Young Scientists, and the Russian Science Support Foundation. 
V.R.M. acknowledges support from the EU Marie Curie Programme, 
Contract No.~MIF1-CT-2006-040816.

\section{Appendix}

Let us assume that for a square pinned lattice, an infinite kink can have a
smooth structure and is directed along the $y$ axis ($x$ and $y$ axes are
directed like in Fig. 1(c)), thus separating two regions of vortex lattices
shifted by $a$ in $y$ direction with respect to each other, as for usual
sine-Gordon systems \cite{Lub}. We also denote kink's length along $x$ as $%
l_{2D}$ ($l_{2D}\gg a$). There are $l_{2D}/a\gg 1$\ vortex rows within the
kink's length and we denote deviations of vortices in the $n$'s row along
the $y$ direction as $u_{n}$. We also restrict ourselves to lattices with
intervortex distances of the order of the penetration depth $\lambda $\ ($%
a\sim 1$ in dimensionless units). In this case, the dominant contribution to
the interaction energy of a given vortex with all other rows of vortices is
provided by harmonics with $m=0$, $\pm 1$, as can be seen from the exponent
in the RHS of Eq. (3). This energy is then given by%
\begin{equation}
E_{n}\simeq \frac{\left( 2\pi \right) ^{2}}{\kappa ^{2}a}\frac{e^{-a}}{%
1-e^{-a}}+\frac{2\pi }{\kappa ^{2}}e^{-2\pi }  \tag{A1}
\end{equation}%
\begin{equation*}
\times \left[ \cos \left( \frac{2\pi (u_{n+1}-u_{n})}{a}\right) +\cos \left( 
\frac{2\pi (u_{n}-u_{n-1})}{a}\right) \right] .
\end{equation*}%
Now we can expand the RHS of Eq. (A1) in terms of $\Delta u_{n}/a$. To find
the energy of the kink, we have to sum the contributions from all the rows
within the kink. But instead it is convenient to switch from summation to
integration with the simultaneous exchange of $\Delta u_{n}/a$ by $\frac{du}{%
dx}$, i.e., to use a continuous limit and introduce smoothly varying
deformation field of the vortex lattice. This method is applicable for
sine-Gordon systems \cite{Lub}, and we repeat the same arguments for the
system at hand. Finally, for the $u$ dependent part of the vortex-vortex
energy of the kink (per length of the kink in $y$ direction), we have:%
\begin{equation}
E_{kink}^{(vv)}=-\frac{\left( 2\pi \right) ^{3}}{\kappa ^{2}a^{2}}e^{-2\pi
}\int \left( \frac{du}{dx}\right) ^{2}dx.  \tag{A2}
\end{equation}%
The derivative $\frac{du}{dx}$ can be estimated as $\frac{du}{dx}\approx 
\frac{a}{l_{2D}}$, and this leads us to the following expression%
\begin{equation}
E_{kink}^{(vv)}\approx -\frac{\left( 2\pi \right) ^{3}}{\kappa ^{2}}e^{-2\pi
}\frac{1}{l_{2D}}.  \tag{A3}
\end{equation}%
There is also a contribution to the kink energy coming from the regular
pinning potential. It appears due to the fact that vortex rows within this
kink are depinned. In the limit of small potential wells, $\sigma \ll a$,
this energy (per length of the kink in $y$ direction) is given by 
\begin{equation}
E_{kink}^{(vp)}\approx U_{0}\frac{l_{2D}}{a^{2}}.  \tag{A4}
\end{equation}%
For the total energy of the kink we thus have%
\begin{equation}
E_{kink}=E_{kink}^{(vv)}+E_{kink}^{(vp)}\approx -\frac{\left( 2\pi \right)
^{3}}{\kappa ^{2}}e^{-2\pi }\frac{1}{l_{2D}}+U_{0}\frac{l_{2D}}{a^{2}} 
\tag{A5}
\end{equation}

If we now try to minimize the kink energy given by Eq. (A5) with respect to
kink's length, we immediately see that $l_{2D}=0$ is the lowest energy
solution. Of course, in reality $l_{2D}$ cannot be shorter than $a$, and $%
l_{2D}=a$ is a true solution. This conclusion is based on the physical fact
that a square lattice of free interacting vortices is unstable in the
absence of pinning, leading to a \textit{negativeness} of "elastic
constant", in contrast to the sine-Gordon and related systems \cite{Lub}.


\begin{references}

\bibitem{1}
M.~Baert, V.V.~Metlushko, R.~Jonckheere, V.V.~Moshchalkov, and Y.~Bruynseraede, 
Phys. Rev. Lett. {\bf 74}, 3269 (1995).

\bibitem{2}
V.V.~Moshchalkov, M.~Baert, V.V.~Metlushko, E.~Rosseel, M.J.~Van~Bael, K.~Temst, R.~Jonckheere, and Y.~Bruynseraede, Phys. Rev. B {\bf 54}, 7385 (1996).

\bibitem{3}
A.T.~Fiory, A.F.~Hebard, and S.~Somekh, Appl. Phys. Lett. {\bf 32}, 73 (1978). 

\bibitem{4}
A.M.~Castellanos, R.~W\"{o}rdenweber, G.~Ockenfuss, A.v.d.~Hart, and K.~Keck, 
Appl. Phys. Lett. {\bf 71}, 962 (1997); 
R. W\"{o}rdenweber, P. Dymashevski, and V.R. Misko, Phys. Rev. B {\bf 69}, 184504 (2004).

\bibitem{5}
J.~Eisenmenger, Z.-P.~Li, W.A.A.~Macedo, and I.K.~Schuller, 
Phys. Rev. Lett. {\bf 94}, 057203 (2005). 

\bibitem{6} 
J.E.~Villegas, E.M.~Gonzalez, M.I.~Montero, I.K.~Schuller, J.L.~Vicent, 
Phys. Rev. B {\bf 68}, 224504 (2003); 
M.I.~Montero, J.J.~Akerman, A.~Varilci, I.K.~Schuller, 
Europhys. Lett. {\bf 63}, 118 (2003). 

\bibitem{7} 
J.E. Villegas, S. Savel'ev, F. Nori, E.M. Gonzalez, J.V. Anguita, R. Garc{\'i}a, 
and J.L. Vicent, Science {\bf 302}, 1188 (2003).

\bibitem{8}
L.~Van~Look, B.Y.~Zhu, R.~Jonckheere, B.R.~Zhao, Z.X.~Zhao, and V.V.~Moshchalkov, 
Phys. Rev. B {\bf 66}, 214511 (2002).

\bibitem{9}
A.V.~Silhanek, S.~Raedts, M.~Lange, and V.V.~Moshchalkov, 
Phys. Rev. B {\bf 67}, 064502 (2003).

\bibitem{10}
M.~Kemmler, C.~G\"{u}rlich, A.~Sterck, H.~P\"{o}hler, N.~Neuhaus,
M.~Siegel, R.~Kleiner, and D.~Koelle, Phys. Rev. Lett. {\bf 97}, 147003 (2006).

\bibitem{11}
A.V. Silhanek, W. Gillijns, V. V. Moshchalkov, B. Y. Zhu, J. Moonens, 
and L. H. A. Leunissen, Appl. Phys. Lett. {\bf 89}, 152507 (2006). 

\bibitem{12}
F.~Nori, Science {\bf 271}, 1373 (1996);
C.~Reichhardt, J.~Groth, C.J.~Olson, S.~Field, and F.~Nori, Phys. Rev. B {\bf 52}, 
10~441 (1995); 
B {\bf 53}, R8898 (1996);
B {\bf 54}, 16~108 (1996);
B {\bf 56}, 14~196 (1997). 

\bibitem{13}
C.~Reichhardt, C.J.~Olson, and F.~Nori, Phys. Rev. B {\bf 57}, 7937 (1998).

\bibitem{14}
C.~Reichhardt, C.J.~Olson, and F.~Nori, Phys. Rev. Lett. {\bf 78}, 2648 (1997); 
Phys. Rev. B {\bf 58}, 6534 (1998). 

\bibitem{15}
V.R.~Misko, S.~Savel'ev, A.L.~Rakhmanov, and F.~Nori, Phys. Rev. Lett. {\bf 96}, 127004 (2006); Phys. Rev. B {\bf 75}, 024509 (2007). 

\bibitem{16}
G.R. Berdiyorov, M.V. Milo\v{s}evi\'{c} and F.M. Peeters, 
Phys. Rev. B {\bf 74}, 174512 (2006). 

\bibitem{17}
G.R. Berdiyorov, M.V. Milo\v{s}evi\'{c} and F.M. Peeters, 
Phys. Rev. B {\bf 76}, 134508 (2007). 

\bibitem{18}
V. Misko, S. Savel'ev, and F. Nori, Phys. Rev. Lett. {\bf 95}, 177007 (2005); 
V.R. Misko, S. Savel'ev, and F. Nori, Phys. Rev. B {\bf 74}, 024522 (2006); 
Physica C {\bf 437-438}, 213 (2006). 

\bibitem{Ukr}  V.N. Rudko, O.N. Shevtsova, and S.V. Shiyanovsky, Fiz. Nizk. Temp. {\bf 22}, 1314 (1996).

\bibitem{Reich2001} 
C. Reichhardt, C. J. Olson, R. T. Scalettar, and G. T. Zimanyi, Phys. Rev. B {\bf 64}, 144509 (2001).

\bibitem{Pogosov} 
W.V. Pogosov, A.L. Rakhmanov, and  V.V. Moshchalkov, Phys. Rev. B {\bf 67}, 014532 (2003).

\bibitem{VVM} 
Q.H. Chen, G. Teniers, B.B. Jin, and V.V. Moshchalkov, Phys. Rev. B {\bf 73}, 014506 (2006).

\bibitem{Reich} 
C. Reichhardt and C.J. Reichhardt, Phys. Rev. B {\bf 76}, 064523 (2007).

\bibitem{Maniv} 
V. Zhuravlev and T. Maniv, Phys. Rev. B {\bf 68}, 174507 (2003).

\bibitem{Golib}
G.R. Berdiyorov, M.V. Milo\v{s}evi\'{c} and F.M. Peeters, 
Physica C {\bf 468}, 809 (2008). 

\bibitem{R1} 
J.W. Reijnders and R.A. Duine, Phys. Rev. Lett. {\bf 93}, 060401 (2004).

\bibitem{R2} 
J.W. Reijnders and R.A. Duine, Phys. Rev. A {\bf 71}, 063607 (2005).

\bibitem{Bigelow} 
H. Pu, L.O. Baksmaty, S. Yi, and N.P. Bigelow, Phys. Rev. Lett. 
{\bf 94}, 190401 (2005).

\bibitem{Cornell} 
S. Tung, V. Schweikhard, and E.A. Cornell, Phys. Rev. Lett. {\bf 97}, 240402 (2006).

\bibitem{French} 
G. Coupier, M. Saint Jean, and C. Guthmann, Phys. Rev. B {\bf 75}, 224103  (2007).

\bibitem{colloids1} 
K. Mangold, P. Leiderer, and C. Bechinger, Phys. Rev. Lett. {\bf 90}, 158302 (2003).

\bibitem{colloids2} 
D. Deb and H.H. von Grunberg, J. Phys.: Condens. Matter {\bf 20}, 245104 (2008).

\bibitem{Zhukov} 
A.A. Zhukov, H. Kupfer, G.K. Perkins, A.D. Caplin, T. Wolf, K.I. Kugel, A.L. Rakhmanov, M.G. Mikheev, V.I. Voronkova, M. Klaser, 
H. Wuhl, Phys. Rev. B {\bf 59}, 11213  (1999).

\bibitem{AL} 
K.I. Kugel, A.L. Rakhmanov, A.A. Zhukov, Physica C {\bf 334}, 203  (2000).

\bibitem{Fisher} 
I.F. Voloshin, A.V. Kalinov, L.M. Fisher, K.I. Kugel, A.L. Rakhmanov, JETP {\bf 84}, 1177 (1997). 

\bibitem{harada}
K.~Harada, O.~Kamimura, H.~Kasai, T.~Matsuda, A.~Tonomura, and V.V.~Moshchalkov, Science {\bf 274}, 1167 (1996). 

\bibitem{togawa}
Y.~Togawa, K.~Harada, T.~Akashi, H.~Kasai, T.~Matsuda, F.~Nori, 
A.~Maeda, and A.~Tonomura, Phys. Rev. Lett. {\bf 95}, 087002 (2005). 

\bibitem{irina07}
I.V.~Grigorieva, W.~Escoffier, V.R. Misko, B.J. Baelus, F.M. Peeters, 
L.Y. Vinnikov and S. Dubonos, Phys. Rev. Lett. {\bf 99} 147003 (2007). 

\bibitem{triangle}
H.J.~Zhao, V.R.~Misko, F.M.~Peeters, S.~Dubonos, V.~Oboznov, 
and I.V.~Grigorieva, Europhys. Lett. {\bf 83} 17008 (2008). 

\bibitem{Lub} 
P.M. Chaikin and T.C. Lubensky, {\it Principles of condensed
matter physics}, Cambridge Univ. Press, Cambridge (1995).

\bibitem{Larkin}
G. Blatter, M.V. Feigel'man, V.B. Geshkenbein, A.I. Larkin, and 
V.M. Vinokur, Rev. Mod. Phys. {\bf 66}, 1125 (1994). 


\end{references}
\end{document}